\documentclass[pre,twocolumn,twoside,showpacs,byrevtex,superscriptaddress]{revtex4}

\usepackage{graphicx,epsfig,verbatim,enumerate}
\usepackage{amssymb,amsmath}
\usepackage{ifthen}
\newboolean{twocolswitch}

\usepackage{color}

\newcommand{\sindex}[1]{}
\newcommand{\nindex}[1]{}

\newcommand{\www}[1]{\url{#1}}

\newcommand{\req}[1]{(\ref{#1})}
\newcommand{\Req}[1]{Eq.~(\ref{#1})}

\newcommand{\tdiff}[2]{\mbox{d} #1/\mbox{d} #2}

\newcommand{\phinode}{\tau}

\newcommand{\kavg}{k_{\rm avg}}

\newcommand{\ka}{k_a}
\newcommand{\kb}{k_b}

\newcommand{\Strig}{S_{\rm trig}}
\newcommand{\Svuln}{S_{\rm vuln}}
\newcommand{\Ptrig}{P_{\rm trig}}

\newcommand{\state}[1]{\sigma_#1}

\newcommand{\infprob}{b}

\setboolean{twocolswitch}{true}

\begin{document}

\title{
  Analysis of a threshold model of social contagion on degree-correlated networks.

}

\author{
\firstname{Peter Sheridan}
\surname{Dodds}
}

\email{peter.dodds@uvm.edu}

\affiliation{Department of Mathematics \& Statistics,
  The University of Vermont,
  Burlington, VT 05401.}

\affiliation{Complex Systems Center
  \& the Vermont Advanced Computing Center,
  The University of Vermont,
  Burlington, VT 05401.}

\author{
\firstname{Joshua L.}
\surname{Payne}
}

\email{joshua.payne@uvm.edu}

\affiliation{Complex Systems Center
  \& the Vermont Advanced Computing Center,
  The University of Vermont,
  Burlington, VT 05401.}

\affiliation{Department of Computer Science
  The University of Vermont,
  Burlington, VT 05401.}

\markboth{Title}
{Author names}

\date{\today}

\begin{abstract}
  We analytically determine when a range of abstract social contagion
models permit global spreading from a single seed on degree-correlated,
undirected random networks.  
We deduce the expected size of the largest
vulnerable component, a network's tinderbox-like critical mass, as
well as the probability that infecting a randomly chosen individual
seed will trigger global spreading.  
In the appropriate limits, our
results naturally reduce to standard ones for models of disease
spreading and to the condition for the existence of a giant component.
Recent advances in the distributed, infinite seed case allow us to
further determine the final size of global spreading events, when they
occur.  
To provide support for our results, we derive exact
expressions for key spreading quantities for a simple yet rich family
of random networks with bimodal degree distributions.

\end{abstract}


\maketitle

Over the last decade, the study of real-world,
complex networks has grown enormously, fueled in no small part
by the advent of readily available, 
large-scale data sets for real systems~\citep{newman2003a,boccaletti2006a}.  
Understanding the coupled dynamics of both the structural evolution
of, and processes on, complex networks remains a fertile area of 
investigation.  Of particular importance is the study of contagion, how entities 
spread through networked 
systems, as exemplified by 
the diffusion of practices,
beliefs, ideas, and emotions in social networks~\citep{coleman1966a,hatfield1993a,fowler2008a}, 
disease contagion in human and animal populations~\citep{murray2002a,anderson1991a}, 
cascading failures in electrical systems~\citep{sachtjen2000a,kinney2005a},
the global spread of computer viruses on the Web~\citep{newman2002d,balthrop2004a},
and the collapse of financial systems~\citep{sornette2003b}.

Our present interest lies in contagion processes where
individuals adopt alternate behaviors through 
imitation of peers.  
We specifically investigate a threshold model of social contagion
on random networks---first proposed and studied by Watts~\citep{watts2002a}---with 
the added complication of arbitrary degree-degree correlations. 
The model's origins lie in the seminal work of
Schelling, who employed a threshold
model for a population on a checkerboard to gain
insight into residential segregation~\citep{schelling1971a,schelling1978a}.
Granovetter~\citep{granovetter1978a} subsequently studied
a mean-field, random mixing threshold model which can
be seen as a natural limiting case of the model we consider here.
Both Schelling and Granovetter's work clearly showed that
global uniformity should not be taken to mean that individuals
have strong or similar preferences, and that small changes
in the distributions of individual preferences could lead
to sharp transitions in the system's macroscopic state.

Placing the threshold model on standard random networks
with arbitrary degree distributions gives rise to
a number of novel behaviors not seen in the Granovetter model.
For example, individuals are now distinguishable, and a single node
changing its state can lead to a complete transition
in the entire system's state~\cite{watts2002a}.
Moreover, by greatly limiting
nodes' knowledge of the complete network, behaviors that would
immediately die out when nodes are aware of all other nodes' states
may now spread globally.  Balkanization enhances innovation.

We consider binary systems with nodes being in one of 
two states, $\state{0}$ and $\state{1}$.
Initially, all nodes are in state $\state{0}$.
Our immediate interest is in determining 
whether or not global spreading is possible
when a single node is switched to state $\state{1}$,
and how this condition is affected by altering the level
of degree-degree correlations.  We then further 
wish to know two key quantities:
(1) the probability global spreading occurs ($\Ptrig$), 
and (2) when it does, to what fraction of
the entire network ($S$).

In finding the probability that global spreading
takes off after switching a single node to state $\state{1}$,
it is enough for us to view the problem as one
of standard percolation~\citep{stauffer1992a,newman2001b}.
(Determining the final extent of spreading requires
a distinct approach but nevertheless still 
capitalizes on the locally branching nature of random
networks~\citep{gleeson2007a,gleeson2008a}.)
This observation follows from several 
well known aspects of random networks.

First, infinite, sparse random networks are locally 
perfectly branching networks, possessing only very
long cycles when sufficiently connected.
Thus, a node can switch to state $\state{1}$ 
in the initial stages of a spreading process (started by a single seed) 
only if a single neighbor switches to $\state{1}$ earlier on.
For global spreading to occur, a network
must have a percolating component of these
easily switched `vulnerable' nodes~\citep{watts2002a}.
(By percolating component, we mean a connected sub-network containing
a non-zero fraction of all nodes.)  
As the adoption of state $\state{1}$ spreads through 
this percolating component of vulnerable nodes,
non-vulnerable nodes requiring two or more neighbors in state $\state{1}$
may also begin to switch.
However, we need focus only on vulnerable nodes
to determine whether global spreading is possible or not.
When a percolating vulnerable component exists,
it may be viewed as a network's critical mass,
and one that is highly susceptible
since it possesses its own critical mass---exactly any one of its own members.

Second, our analytic treatment via generating functions
is limited to 
the description of finite network components,
which indirectly allows us
to describe some aspects of infinite
components.  Since finite components
are pure branching structures
(i.e., they contain no cycles of any length),
all nodes can only
switch to $\state{1}$ due to the
conversion of a single neighbor,
even in the long run.

We structure the paper as follows.
After defining the model
fully in Section~\ref{sec:tccn.modeldefn},
we detail a series of analytic results
in Section~\ref{sec:tccn.theory},
concerning the probability and 
size of macroscopic spreading events.
We confirm, via 
further analysis and simulations, a number of our calculations 
for a simple network containing
two kinds of nodes in Section~\ref{sec:tccn.numerics}.
We offer some concluding remarks
in Section~\ref{sec:tccn.conclusion}.

\section{Model definition}
\label{sec:tccn.modeldefn}

Each node is initially assigned a `response function'
which we take here to be a step function.
In effect, each node is given a fixed
threshold $\phinode$ sampled from a distribution $P_{\rm threshold}(\phinode)$.
Node states update in synchrony
at times $t=0$, 1, 2, \ldots
Each node observes the fraction of its neighbors in
state $\state{1}$, and switches to $\state{1}$ if this fraction
meets or exceeds its threshold, $\phinode$.
Once a node switches to state $\state{1}$, it remains
in state $\state{1}$ permanently
(akin to the SI model for disease spreading~\citep{murray2002a}).
Asynchronous updating gives the same
results for monotonically increasing response functions.

We define the structure of our networks through edge probabilities,
following Newman~\citep{newman2002a,newman2003e}.
In studying a range of structural aspects of degree-correlated networks 
as well as dynamic phenomena on them (especially contagion processes),
it is mathematically convenient to use $e_{jk}$,
the probability that a randomly chosen edge
connects a node with degree $j+1$ to one with degree $k+1$,
rather than $j$ and $k$.
The quantity $e_{jk}$ then refers directly to 
the number of other edges emanating from
the nodes an edge connects.
Normalization is uncomplicated, requiring that 
$
\sum_{j=0}^{\infty}
\sum_{k=0}^{\infty}
e_{jk}
= 1
$.

While this definition of $e_{jk}$ is clear for directed networks,
we encounter some subtleties in attempting to accommodate 
both undirected and directed networks with a single notation
(even though many expressions will turn out to be the same).
For undirected networks, which are our primary focus here, we evenly divide
the probability that a randomly selected edge (now directionless) 
connects nodes with degree $j+1$ and $k+1$ between 
the quantities $e_{jk}$ and $e_{kj}$.  The chance of 
a randomly selected edge connecting degree $j+1$ and $k+1$
nodes is then $e_{jk} + e_{kj} = 2e_{jk}$, and the matrix
formed by the $e_{jk}$ is symmetric.  We are thus
effectively retaining a ghost of a directed network, as each link must 
have a designated first and second node.
(For directed networks, $e_{jk}$ need of course not be symmetric,
and other complications are possible concerning the correlations between
an individual node's in-degree and out-degree.)

We also have 
the important quantity
$
q_k
$
which is the probability 
that in randomly choosing an edge,
and then randomly choosing one end
of that edge, we arrive at a node
of degree $k+1$ (equivalently, the 
node has $k$ emanating edges).
For undirected networks, we have
the derivation
$
q_k
=
\frac{1}{2}\sum_{j=0}^{\infty}
e_{jk}
+
\frac{1}{2}\sum_{j=0}^{\infty}
e_{kj}
=
\sum_{j=0}^{\infty}
e_{jk}.
$
(The same end expression holds for directed networks,
where we must follow the edge's direction.)

The Pearson correlation coefficient $r$ for degree pairs
gives us a measure of assortativity, and is given by
\begin{equation} 
  \label{eq:tccn.assort}
  r = \frac{1}{\sigma_q^2}
  \sum_{j=0}^{\infty}
  \sum_{k=0}^{\infty}
  {jk(e_{jk} - q_jq_k)}
\end{equation}
where $\sigma_q^2 = \sum_k{k^2q_k} - [\sum_k{kq_k}]^2$
is the variance in the number of emanating edges
from a node arrived at by a random edge.
Note that the choice of $e_{jk}$ almost prescribes
the form of the resulting network's degree distribution, $p_k$.
The one piece of information missing is the abundance
of nodes with no connections, $p_0$, which we
must define independently.  
A link between
the $e_{jk}$ and $p_k$ for undirected networks
follows from the observation that $q_k$ is
readily determined from the degree distribution:
a randomly chosen edge leads to a degree $k+1$ 
node with probability $q_k = (k+1)p_{k+1}/\kavg$ 
where $\kavg = \sum_{k=0}^\infty k p_k$ is the average degree.
We therefore have the connection
$
q_k 
=
\sum_{j=0}^\infty e_{jk}
=
(k+1)p_{k+1}/\kavg.
$
Now, in isolating the $p_k$, we do not need to know $\kavg$.
It is enough to know $p_k \propto q_k/(k+1)$ for $k>0$, 
since we can find the normalization constant (which is in fact $\kavg$)
by requiring $\sum_{k=1}^\infty p_k = 1 - p_0$.

Finding the probability of triggering a global
spreading event reduces to standard percolation
once we find the probability that an individual
node is vulnerable.  We allow this probability
to be a function of node degree $k$, using
the notation $\infprob_{k1}$ (more generally, we write the probability
that a node of degree $k$ switches to state $\state{1}$
given $l$ contacts in state $\state{1}$ as $\infprob_{kl}$).
We first determine
whether nodes are vulnerable or not, removing
them from the network in the latter case.
Finally, if global spreading occurs on the resulting reduced network,
global spreading must occur on the original network.

\section{Analytic results}
\label{sec:tccn.theory}

Building on the work of Newman~\citep{newman2002a}, we find
closed form expressions for several probability generating functions
related to component sizes.  
The key probability we need to 
characterize is $f_{n,j}$, which is the probability that
an edge emanating from a degree $j+1$ node leads to
a finite vulnerable subcomponent of size $n = 0,1,2,\ldots$.
Writing the marginal generating function for $f_{n,j}$ as
$F_j(x;\vec{\infprob}_1) = \sum_{n=0}^\infty f_{n,j} x^n$, we have the 
following recursive relationship:
\begin{align}
  \label{eq:tccn.F_jF_k}
  \nonumber
  F_j(x;\vec{\infprob}_1)
  & =
  x^{0}
  \sum_{k=0}^{\infty} 
  \frac{e_{jk}}{q_j}
  ( 1 - \infprob_{k+1,1} )
  \\
  & +
  x
  \sum_{k=0}^{\infty} 
  \frac{e_{jk}}{q_j}
  \infprob_{k+1,1}
  \left[
    F_k(x;\vec{\infprob}_1)
  \right]^k,
\end{align}
where $j=0,1,2, \ldots$.
In both terms, we have the quantity
$e_{jk}/q_j$
which represents the normalized probability
that an edge from a degree node $j+1$ leads
to a degree $k+1$ node.
For undirected networks, we obtain this probability
as
$
(e_{jk} + e_{kj})/(\sum_{k=0}^{\infty} e_{jk} + e_{kj}) = 2e_{jk}/2q_j = e_{jk}/q_j.
$
The first term in
\Req{eq:tccn.F_jF_k}
records the probability
that an edge from a degree $j+1$ node leads immediately
to a non-vulnerable node, and hence a 
vulnerable subcomponent of size $0$.

The second term involves a composition of 
generating functions.  The generating function
$F_k$ is the argument of 
\begin{equation}
  \label{eq:tccn.gf}
  \sum_{k=0}^{\infty} 
  \frac{e_{jk}}{q_j}
  \infprob_{k+1,1}
  x^k,
\end{equation}
which is itself the generating function
for the probability that an edge
from a degree $j+1$ node leads to a vulnerable
node with $k$ emanating edges~\cite{footnote:tccn.composition}.
Finally, the $x$ leading the second term in \Req{eq:tccn.F_jF_k}
accounts for the $k+1$ degree node itself.

Our task is now to find critical points 
indicating the onset of a giant component
as we vary network structure by altering the $e_{jk}$ and $\infprob_{k1}$.
We compute the average size of a finite vulnerable component
found by following an edge from a degree $j+1$ node,
given by $F_j'(1;\vec{\infprob}_1)$.  
Differentiating \Req{eq:tccn.F_jF_k},
setting $x=1$,
and substituting $q_j = \sum_{k=0}^{\infty} e_{jk}$,
we have
\begin{equation}
  \label{eq:tccn.F_j'1}
  q_j F_{j}'(1;\vec{\infprob}_1)
  =
  \sum_{k=0}^{\infty}
  e_{jk}
  \infprob_{k+1,1}
  + 
  \sum_{k=0}^{\infty}
  k e_{jk}
  \infprob_{k+1,1}
  F_k'(1;\vec{\infprob}_1).
\end{equation}
(We have used the fact that $F_k(1;\vec{\infprob}_1)=1$ for
networks without a giant component, and all
components contribute to the generating function of $f_{n,k}$.)
Rearranging \Req{eq:tccn.F_j'1}, we have a linear system
\begin{equation}
  \label{eq:tccn.F_j'2}
  \sum_{k=0}^\infty
  \left(
    \delta_{jk} q_k 
    -
    k \infprob_{k+1,1} e_{jk}
  \right)
  F_k'(1;\vec{\infprob}_1)
  =
  \sum_{k=0}^{\infty}
  e_{jk}
  \infprob_{k+1,1}
\end{equation}
which we write as
\begin{equation}
  \label{eq:tccn.F_j'3}
  \mathbf{A}_{\mathbf{E},\vec{\infprob}_1 }
  \vec{F}'(1;\vec{\infprob}_1)
  =
  \mathbf{E} \vec{\infprob}_1
\end{equation}
where 
\begin{align}
  \label{eq:tccn.F_j'defs1}
  \left[ \mathbf{A}_{\mathbf{E},\vec{\infprob}_1 } \right]_{j+1,k+1}
   & = 
  \delta_{jk} q_{k} 
  -
  k \infprob_{k+1,1} e_{jk},
  \\
  \label{eq:tccn.F_j'defs2}
  \left[ \vec{F}'(1;\vec{\infprob}_1)  \right]_{k+1}
   & = 
  F_{k}'(1;\vec{\infprob}_1),
  \\
  \label{eq:tccn.F_j'defs3}
  \left[ \mathbf{E} \right]_{j+1,k+1}
   & = 
  e_{jk},
  \\
  \label{eq:tccn.F_j'defs4}
  \mbox{and} \  \left[ \vec{\infprob}_1 \right]_{k+1}
   & =
  \infprob_{k+1,1}
\end{align}
for $j$, $k$ = 0, 1, 2, 3, \ldots.
A solution exists when $\mathbf{A}_{\mathbf{E},\vec{\infprob}_1 }$
is invertible, i.e., its determinant is non-zero.  We then have
\begin{equation}
  \label{eq:tccn.F_j'soln}
  \vec{F}'(1;\vec{\infprob}_1)
  =
  \mathbf{A}_{\mathbf{E},\vec{\infprob}_1 }^{-1}
  \, \mathbf{E} \vec{\infprob}_1.
\end{equation}
Our condition for the onset of global spreading is therefore
\begin{equation}
  \label{eq:tccn.F_j'cond}
  \left| \mathbf{A}_{\mathbf{E},\vec{\infprob}_1 } \right| 
  = 0.
\end{equation}
with $\mathbf{A}_{\mathbf{E},\vec{\infprob}_1 }$ defined in \Req{eq:tccn.F_j'defs1}.
We note that for networks where one or more degrees are not present
(which we will encounter in our later specific examples),
we omit the rows and columns of the above matrices corresponding
to those degrees.  For simplicity, we present our 
general results for networks assuming all degrees are represented,
observing that adjustments to specific sets of degrees are straightforward.

For uncorrelated networks, upon substituting $e_{jk}=q_j q_k$,
the above collapses to the known condition
\begin{equation}
  \label{eq:tccn.uncorrcond}
  \sum_{k=0}^{\infty} \infprob_{k1} (k-1) \frac{k  p_k}{\kavg} = 1,
\end{equation}
as expected~\citep{watts2002a}.  
Moreover, $F_{k}'(1;\vec{\infprob}_1)$ is seen to be independent
of $k$, since the first node of an edge is now unrelated
to the other node and hence also the subcomponent it leads to.
We provide details for these calculations in the Appendix.

Returning to the general case of degree-correlated networks,
we see that when $\infprob_{k1}=\infprob$, a constant for all $k$,
we have a disease-like contagion process.  Furthermore, when
$\infprob=1$, all nodes are vulnerable, and we have found
the condition for a giant component, equivalent to
that obtained by Newman~\citep{newman2002a}.  (Note that when $\infprob_{k1}=\infprob$,
we have $\mathbf{E} \vec{\infprob}_1 = \infprob \vec{q}$ since
$[\mathbf{E} \vec{\infprob}_1]_{j+1} = \sum_{k=0}^{\infty} e_{jk} \infprob_{k+1,1}
= \infprob \sum_{k=0}^{\infty} e_{jk} = \infprob q_{j+1}$.)

We next consider two probability distributions
pertaining to component size:
(1) $g_{n}$, the probability that a randomly chosen node
belongs to a vulnerable component of size $n$,
and (2) $h_{n}$, the probability that a randomly chosen
node belongs or is adjacent to a vulnerable component
of size $n$.  Knowing $g_n$ helps us find the
size of the largest vulnerable component, whose presence
or absence dictates whether or not global cascades are possible
for infinite random networks.
The second probability $h_n$ will aid us in determining
the probability of triggering a global cascade.  
The triggering node, which is exogenously switched to state $\state{1}$
may be either vulnerable and part of the largest vulnerable component,
or non-vulnerable and connected to one or more nodes in
the largest vulnerable component.  
For the  standard giant component case where $\infprob_{k1}=1$, 
we have $g_n=h_n$;
otherwise, these distributions are likely distinct.

As for the $f_{k,n}$, we find closed form
expressions for the generating functions
associated with $g_n$ and $h_n$.
The generating function for $g_{n}$ satisfies the 
following relationship:
\begin{align}
  \label{eq:tccn.Gx}
  \nonumber
  G(x;\vec{\infprob}_1)
  = & 
  \, x^0
  \sum_{k=0}^{\infty}
  p_k(1-\infprob_{k1})
  +
  x^1
  p_0
  \infprob_{01}
  \\
  & + 
  x
  \sum_{k=1}^\infty
  p_k
  \infprob_{k1}
  \left[
    F_{k-1}(x;\vec{\infprob}_1)
  \right]^k.
\end{align}
The $x^0$ term carries the probability
that a randomly selected node will not
be in state $\state{1}$; 
the second term accounts for
the randomly chosen node being vulnerable but
having degree 0;
and the third term again uses the composition
rule for sums of random variables of random sizes~\cite{footnote:tccn.composition}.
The generating function $\sum_{k=1}^{\infty} p_k \infprob_{k1} x^k$
corresponds to the probability distribution for a randomly
chosen node to have degree $k$ and be vulnerable.
Note that the argument $F_{k-1}$ appears in the last term
rather than $F_{k}$ because $F_{k-1}$ by definition
corresponded to a degree $k$ node.

The generating function for the triggering distribution
satisfies a simplified version of \Req{eq:tccn.Gx}:
\begin{equation}
  \label{eq:tccn.Hxbeta}
  H(x;\vec{\infprob}_1)
  = 
  x^1
  p_0
  +
  x
  \sum_{k=1}^\infty
  p_k
  \left[
    F_{k-1}(x;\vec{\infprob}_1)
  \right]^k.
\end{equation}
For the triggering distribution, the first
node is now always switched to state $\state{1}$,
regardless of whether it is itself vulnerable
or not.  The $x^1$ term accounts for this initial node
having degree 0 and hence being unable to trigger
any other node.  If the initial node has at least
one neighbor, then spreading may occur and 
we can make use of the basic degree 
generating function $\sum_{k=1}^{\infty} p_k x^k$
combined with the generating function $F_{k-1}(x;\vec{\infprob}_1)$
for finite vulnerable subcomponent size.
As per our previous examples, the preceding factor $x$ in the second 
term accounts for the initial node.
In effect, $\infprob_{k1}=1$ for the triggering node
as it is always forced to be switched
to state $\state{1}$, and indeed setting $\infprob_{k1}=1$
in \Req{eq:tccn.Gx} directly yields \Req{eq:tccn.Hxbeta}.

Now, the fraction of nodes in the largest vulnerable component
is given by $\Svuln = 1 - G(1;\vec{\infprob}_1)$, since $G(1;\vec{\infprob}_1)$
can be seen as the probability that a random node is part
of a finite vulnerable component (including one of size 0).
Setting $x=1$ in \Req{eq:tccn.Gx}, we have
\begin{align}
  \label{eq:tccn.G1}
  \nonumber
  \Svuln 
  = & 
  1 -
  G(1;\vec{\infprob}_1) \\
  = & 
  \sum_{k=0}^{\infty}
  p_k\infprob_{k1} 
  -
  p_0
  \infprob_{01}
  -
  \sum_{k=1}^\infty
  p_k
  \infprob_{k1}
  \left[
    F_{k-1}(1;\vec{\infprob}_1)
  \right]^k.
\end{align}
In the same fashion as for $\Svuln$, 
the probability of triggering a cascade can be determined
using $H(x;\vec{\infprob}_1)$:
\begin{align}
  \label{eq:tccn.H1}
  \nonumber
  \Strig 
  = & 
  1 -
  H(1;\vec{\infprob}_1) \\
  = &
  1
  -
  \sum_{k=0}^\infty
  p_k
  \left[
    F_{k-1}(1;\vec{\infprob}_1)
  \right]^k.
\end{align}
The size of the triggering
component can also be obtained by first making the observation 
that an initially switched node of degree $k$ 
triggers a cascade with probability
\begin{equation}
  \label{eq:tccn.Strigk}
  \Strig^{(k)}
  = 
  1 -
  \left[
    F_{k-1}(1;\vec{\infprob}_1)
  \right]^k.
\end{equation}
This is the probability that
at least one edge from a degree $k$ node leads
to a giant component of vulnerable nodes. 
We then have
$
  \Strig
  = 
  \sum_{k=0}^{\infty}
  p_k
  \Strig^{(k)},
$
which is in agreement with
\Req{eq:tccn.H1}.

Both $\Svuln$ and $\Strig$
depend on $F_{k}(1;\vec{\infprob}_1)$,
for which we obtain a potentially infinite set of 
closed-form, coupled, nonlinear recursive expressions
from \Req{eq:tccn.F_jF_k}:
\begin{equation}
  \label{eq:tccn.F_jF_kx=1}
  F_j(1;\vec{\infprob}_1)
  =
  \sum_{k=0}^{\infty} 
  \frac{e_{jk}}{q_j}
  (1-\infprob_{k+1,1})
  +
  \sum_{k=0}^{\infty} 
  \frac{e_{jk}}{q_j}
  \infprob_{k+1,1}
  \left[
    F_k(1;\vec{\infprob}_1\,)
  \right]^k
\end{equation}
where we have substituted
$q_j = \sum_{k=0}^{\infty} e_{jk}$.
Solving \Req{eq:tccn.F_jF_kx=1} will almost
always involve numerical techniques
(though in Section~\ref{sec:tccn.numerics}, we examine
a case that has analytic solutions).
The uncorrelated, pure random network version
for the giant component ($\infprob_{k1}=1$)
may serve as some inspiration.  There, we
have no dependence on the degree of
the initial node and the problem reduces
to solving $F(1,\vec{1}) = \sum_{k=0}^{\infty} q_k 
[F(1,\vec{1}) ]^k$.  Both $F(1,\vec{1})=0$ and $F(1,\vec{1})=1$
are solutions and any initial estimate in between
leads to, upon iteration, an intermediate solution,
if one exists.  
Similarly, iteration
of \Req{eq:tccn.F_jF_kx=1}
should generally reach the appropriate fixed point solution.

We can also and more easily determine the average size
of all vulnerable components (applicable when no giant
vulnerable component is present since we again use
$F_k(1;\vec{\infprob}_1)=1$):
\begin{equation}
  \label{eq:tccn.G'(1)soln}
  G'(1,\vec{\infprob}_1) 
  = 
  \vec{p}^{\rm T} \vec{\infprob}_1
  +
  \kavg
  (
    \mathbf{E} \, \vec{\infprob}_1
  )^{\rm T}
  \mathbf{A}_{\mathbf{E},\vec{\infprob}_1}^{-1}
   \, (\mathbf{E} \, \vec{\infprob}_1),
\end{equation}
where $[\vec{p}]_k = p_k$.
If we allow spreading to start from any node
by forcing the first node to be in state $\state{1}$,
then the average size is instead given by
\begin{equation}
  \label{eq:tccn.H'(1)soln}
  H'(1,\vec{\infprob}_1) 
  = 
  1
  +
  \kavg
  (
    \mathbf{E} \, \vec{1}
  )^{\rm T}
  \mathbf{A}_{\mathbf{E},\vec{\infprob}_1}^{-1}
   \, (\mathbf{E} \, \vec{\infprob}_1).
\end{equation}

We complete our analysis with a description
of the expected final size of a global spreading event,
when it occurs.  
We use the results of recent work 
by Gleeson and Calahane~\citep{gleeson2007a},
and subsequently Gleeson~\citep{gleeson2008a},
who have explicated an elegant, general solution to 
the distributed, infinite seed case for a variety
of spreading models on a wide range of random networks.

Their key observation is that a node $i$ (degree $k_i$) switching
to state $\state{1}$
at time step $n$ can only be due to previous switching in nodes 
within $n$ steps of $i$ (including, trivially, $i$ itself).
Furthermore, this neighborhood of $i$ must
be a pure branching network, i.e., a tree rooted at $i$.
At time step $n=1$, only $i$'s immediate neighbors
influence $i$, and these are in state $\state{1}$ with
probability $\phi_0$.  We write the probability that
one of $i$'s edges connects to a node in state $\state{1}$ 
at time step $n+1$ as $\theta_{k_i,n}$.  
At time step $n=2$, the effect of 
next neighbors is felt, and $i$'s degree $k$ neighbors are now
in state $\state{1}$ with probability $\theta_{k,1}$ which
in turn depends generally on $\theta_{k',0}=\phi_0$, $k' \ge 1$.
Allowing $n$ to increase, we are lead to a recursive
expression for the $\theta_{k,n}$ as follows.
The expected size
of a global spreading event given a fraction
$\phi_0$ of nodes active at time $t=0$ is
$\phi_\infty$, which is obtained in the $n \rightarrow \infty$
limit of the following equations~\citep{gleeson2008a}:
\begin{equation}
  \label{eq:tccn.gleeson_corr_phi}
  \phi_{n+1}
  =
  \phi_0
  +
  (1-\phi_0)
  \sum_{k=0}^{\infty}
  p_k
  \sum_{i=0}^{k}
  \binom{k}{i}
  \theta_{k,n}^{\, i}
  (1-\theta_{k,n})^{k-i}
  \infprob_{ki}
\end{equation}
where
\begin{eqnarray}
  \label{eq:tccn.gleeson_corr_r}
  \lefteqn{
    \theta_{j,n+1}
    =
    \phi_0 + 
    (1-\phi_0) \times
  } & &  \\
  & & 
  \sum_{k=1}^{\infty}
  \frac{e_{j-1,k-1}}{q_{j-1}}
  \sum_{i=0}^{k-1}
  \binom{k-1}{i}
  \theta_{k,n}^{\, i}
  (1-\theta_{k,n})^{k-1-i}
  \infprob_{ki}
  \nonumber
\end{eqnarray}
and, again, $\theta_{k,0} = \phi_0$ and
$\infprob_{ki}$ is the probability that a degree
$k$ node switches to state $\state{1}$ if
exactly $i$ of its neighbors have switched.

We thus observe a symmetry between
the explanations for when global spreading 
may occur from a single seed and if so, to what extent.
The former concerns the progress of a spreading
event as it moves outward from a single seed
through a random branching network,
and the latter hinges on how switching converges
on a central node, again traversing a branching
network, but in the reverse direction
(this process is similarly described in~\citep{diekmann1998a}
for a particular disease spreading model involving repeated
contacts).

Note that $\phi_n$ must always increase or stay the same,
since nodes never switch off, and must therefore approach
a limit as $n \rightarrow \infty$.  This implies that node $i$ 
feels the effect of initial activations within a finite
number of steps, and indeed we see that the approach
to $\phi_\infty$ is rapid.  
Whether or not a small seed
takes off can be determined from 
\Req{eq:tccn.gleeson_corr_r} 
by examining the matrix
$\tdiff{\theta_{j,n+1}}{\theta_{k,n}}$ evaluated
at $\theta_{k,n}=0$~\cite{gleeson2008a}, and yields
a condition equivalent to our requirement $|\mathbf{A}| = 0$
(\Req{eq:tccn.F_j'cond}).

While these equations are derived with the assumption
of an infinite seed (if an arbitrarily small one, fractionwise), 
they nevertheless 
perform extremely well in the limit $\phi_0 \rightarrow 0$.
Even so, they cannot capture the salience
of a global vulnerable component for the single seed case, 
which must be present for spreading to succeed.
By comparison, for the infinite seed model, spreading always occurs
if the cascade condition is met.
However, as noted in~\citep{gleeson2008a},
by separately determining $\Strig$, we can
account for the major features of the single seed model:
the probability of generating a global spreading event, $\Strig$, 
and its expected size $\phi_\infty$.
We note that in~\citep{gleeson2008a}, an expression
for $\Strig$ for uncorrelated random networks was obtained
using $\theta_\infty$; here, we have generalized $\Strig$
to uncorrelated networks directly from the standard
generating function approach.

\section{Application to a simple family of random networks}
\label{sec:tccn.numerics}

For a tractable test case, we examine
a family of infinite random networks with nodes having 
either of
two degrees $\ka$ and $\kb$ where
$\ka <  \kb$.  We consider the general class
of threshold profiles $\vec{\infprob}_1$ such
that degree $\ka$ nodes are vulnerable and
degree $\kb$ nodes are not.
We are able to obtain expressions for the 
size of the vulnerable and triggering components
as a function of assortativity,
and we make comparisons
with simulations for the $(\ka,\kb)=(3,4)$ case.

We wish to consider the
full range of assortativity and this constrains the
relative probabilities of the two degrees
in the following way.  We specifically need
\begin{equation}
  \label{eq:tccn.negasst}
  p_{\ka} = 
  \frac{\kb}{\ka}
  p_{\kb},
\end{equation}
since when $r=-1$, degree $\ka$ nodes
connect only to degree $\kb$ nodes,
and so there must be relatively fewer of the latter.
We therefore have
\begin{equation}
  \label{eq:tccn.negasst2}
  p_{\ka} = 
  \frac{\kb}{\ka + \kb}, \
  p_{\kb} = 
  \frac{\ka}{\ka + \kb}, \
  \mbox{and} \
  p_{k} = 0 \ \mbox{otherwise.}
\end{equation}
Values of $r$ other than -1 add no further constraints.
The corresponding $\mathbf{E}$ matrix 
is independent of $\ka$ and $\kb$
and has the form
\begin{equation}
  \label{eq:tccn.negasst3}
  \mathbf{E}
  =
  \frac{1}{4}
  \left[
    \begin{array}{ll}
      1+r & \ 1-r \\
      1-r & \ 1+r \\
    \end{array}
  \right],
\end{equation}
where we have left out irrelevant rows and columns of zeros.
The condition for a giant component
given by 
Eqs.~\req{eq:tccn.F_j'defs1}
and
\req{eq:tccn.F_j'cond}
is then
\begin{eqnarray}
  \label{eq:tccn.Areq}
  0 = |\mathbf{A}_{\mathbf{E},\vec{\infprob}_1 }|
  & =
  \frac{1}{4}
  \left|
    \begin{array}{ll}
      1-\frac{1}{2}(\ka-1)(1+r) & \ 0 \\
      \frac{1}{2}(\ka - 1)(r-1) & \ 1 \\
    \end{array}
  \right|
  \\ \nonumber
  & = \frac{1}{2}-\frac{1}{4}(\ka-1)(1+r).
\end{eqnarray}
We therefore have a phase transition when
\begin{equation}
  \label{eq:tccn.rphasetransition}
  r = -1 + \frac{2}{\ka - 1},
\end{equation}
with a giant component of vulnerable nodes
arising and then growing as assortativity increases.
(Note that when no giant component exists,
$|\mathbf{A}|$ is positive.)  For the case $\ka=3$,
the phase transition occurs at $r=0$,
and as $\ka$ increases, the location
of the phase transition moves to more
negative values of $r$.

We next determine the size of the vulnerable
and triggering components.
We find the probabilities of 
reaching a finite component from
degree $\ka$ and $\kb$ nodes via \Req{eq:tccn.F_jF_kx=1}:
\begin{align}
  \label{eq:tccn.F_jF_kx=1specific}
  F_{\ka-1}(1;\vec{\infprob}_1\,)
  & =
  \frac{1-r}{2}
  +
  \frac{1+r}{2}
  \left[ F_{\ka-1}(1;\vec{\infprob}_1\,) \right]^{\ka-1},
  \\
  F_{\kb-1}(1;\vec{\infprob}_1\,)
  & =
  \frac{1+r}{2}
  +
  \frac{1-r}{2}
  \left[ F_{\ka-1}(1;\vec{\infprob}_1\,) \right]^{\ka-1}.
\end{align}
Only the first equation need be solved and is in general
a $\ka-1$ degree polynomial in $F_{\ka-1}(1;\vec{\infprob}_1\,)$.
We now specialize our results for 
the case $\ka=3$.  
We have a quadratic equation from which we obtain
\begin{equation}
  \label{eq:tccn.Fsolns}
  F_{2}(1;\vec{\infprob}_1\,) = 
  \left\{
    \begin{array}{cc}
      1 & \mbox{for $r \le 0$}, \\
      \frac{1-r}{1+r} & \mbox{for $r \ge 0$}.
    \end{array}
  \right.
\end{equation}
and
\begin{equation}
  \label{eq:tccn.Fsolns2}
  F_{\kb-1}(1;\vec{\infprob}_1\,) = 
  \left\{
    \begin{array}{cc}
      1 & \mbox{for $r \le 0$}, \\
      \frac{1+r}{2}
      \left[
        1 +
        \left(
        \frac{1-r}{1+r} 
        \right)^{3}
      \right]
      & \mbox{for $r \ge 0$}.
    \end{array}
  \right.
\end{equation}
Note that in finding $F_{2}(1;\vec{\infprob}_1\,)$,
we have had to make a choice between the two roots
of the quadratic given in~\Req{eq:tccn.Fsolns}.
First, we have only one realistic solution in the interval $[0,1]$
for $r<0$, namely $F_{2}(1;\vec{\infprob}_1\,)=1$
(the other solution, $(1-r)/(1+r)$, exceeds 1 for $r<0$).
Second, in considering the limit $r=1$ where degree 3 nodes
are connected only to other degree 3 nodes forming
a giant component, we require that the probability
of reaching a finite component must drop to 0.
We therefore see that
$(1-r)/(1+r)$ must be the correct choice since
then $F_{2}(1;\vec{\infprob}_1\,)=0$ at $r=1$.
We further observe that the two roots cross when $r=0$, and for $r>0$,
the root $(1-r)/(1+r)$ is less than the root at unity.
Thus, by an expectation of continuity, $(1-r)/(1+r)$  must be
the solution for all of $r \ge 0$.

We next determine the size
of the vulnerable component and triggering probabilities
as functions of $r$
using Eqs.~\req{eq:tccn.G1}, \req{eq:tccn.H1}, and~\req{eq:tccn.Strigk}.
For $r < 0$,
we have $\Svuln = \Strig = 0$,
and for $r \ge 0$,
\begin{equation}
  \label{eq:tccn.G1specific}
  \Svuln 
  =
  \frac{\kb}{\kb+3}
  \left[
    1
    -
    \left(
      \frac{1-r}{1+r}
    \right)^3
  \right],
\end{equation}
\begin{align}
  \label{eq:tccn.H1specific}
  \Strig^{(3)}
  = &
  1 
  - 
  \left(
    \frac{1-r}{1+r}
  \right)^3, \\ \nonumber
  \Strig^{(k_b)}
  = &
  1 
  - 
  \left(
    \frac{1+r}{2}
  \right)^{\kb}
  \left[
    1 +
    \left(
      \frac{1-r}{1+r} 
    \right)^{3}
  \right]^{k_b},
\end{align}
and
\begin{align}
  \label{eq:tccn.H1specific2}
  \Strig 
  = &
  1 
  - 
  \frac{\kb}{\kb+3}
  \left(
    \frac{1-r}{1+r}
  \right)^3 \\ \nonumber
  & -
  \frac{3}{\kb+3}
  \left(
    \frac{1+r}{2}
  \right)^{\kb}
  \left[
    1 +
    \left(
      \frac{1-r}{1+r} 
    \right)^{3}
  \right]^{k_b}.
\end{align}

\begin{figure}[tbp]
  \centering
  \includegraphics[width=\columnwidth]{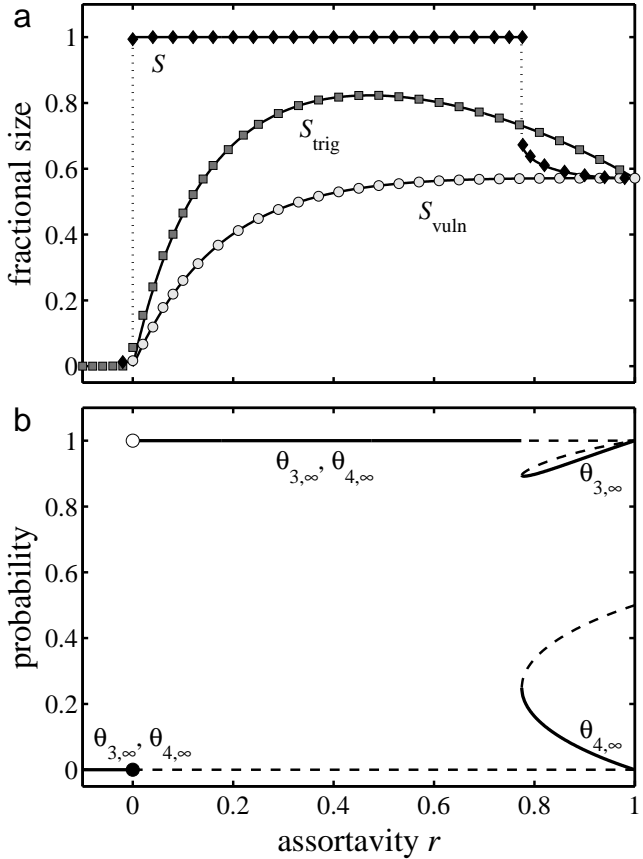}
  \caption{
    \protect
    \textbf{(a)} 
    Three key spreading measures as a function of assortativity $r$
    for networks with $P_k = \frac{4}{7} \delta_{k3} + \frac{3}{7} \delta_{k4}$.
    Solid lines indicate theoretical curves and symbols represent
    measurements from simulations.
    The three curves are (1) the fractional size of the largest vulnerable component,
    $\Svuln$, (\Req{eq:tccn.G1specific}, circles); 
    (2) the fractional size of the largest triggering component,
    $\Strig$, (\Req{eq:tccn.H1specific2}, squares); 
    and
    (3) the fractional size of global spreading events $S$
    (\Req{eq:tccn.gleeson_corr_phi}, diamonds).
    The upper discontinuous phase transition in $S$ 
    occurs at $r = r_c^{\rm upper} = 0.77485 \pm 0.00001$.
    The data shown were obtained from networks with $N$=0.98$\times 10^5$ nodes
    (correlations were generated using a shuffling algorithm described in~\cite{payne2008a})
    with initial seeds placed at each 
    node of 10 sample networks for a total of 9.8$\times 10^5$ samples per each value of $r$.
    \textbf{(b)}
    Solutions for $\theta_{3,\infty}$ and $\theta_{4,\infty}$,
    the probability that edges leading to degree 3 and 4 nodes
    will be from nodes in state $\state{1}$ in the long time limit
    (see Eqs.~\req{eq:tccn.Sr3/4} and~\req{eq:tccn.Sr3/4_b}.)
    Solid and dashed lines indicate stable and unstable ones
    respectively.
    When $r$ reaches $r_c^{\rm upper}$, a repeated solution
    appears for $(\theta_{3,\infty},\theta_{4,\infty})$,
    distinct from $(1,1)$, giving rise to the discontinuous
    upper phase transition in $S$.  
  }
  \label{fig:tccn.modelfit}
\end{figure}

As shown in Fig.~\ref{fig:tccn.modelfit}(a), we obtain
excellent agreement between equations and the output 
of simulations for $k_b=4$ on networks with $N=0.98 \times 10^5$ nodes
and $\phinode=1/3$.  
We used this particular value of $N$ since
it is the nearest number to $10^5$ that is divisible by $7$ 
[required by Eqs.~\req{eq:tccn.negasst} and \req{eq:tccn.negasst2}]
and, for convenience, also a multiple of 100.
As was observed for the original threshold model on random
networks~\citep{watts2002a}, we find
that well-connected networks near the transition to a non-percolating largest vulnerable
component are of a
`robust-yet-fragile' nature~\citep{carlson2000a}.
We see in Fig.~\ref{fig:tccn.modelfit}(a) that
if assortativity $r$ is just slightly positive, and if the initial seed 
is in the small triggering component present, then spreading reaches
the entire network.  
These otherwise highly resilient networks have
an Achilles heel that leads to a complete transition in individual
node states.  Furthermore, finite size effects may be significant:
for the networks with $N=9800$ nodes,
our simulations show that global spreading is possible
for $r$ as low as $-0.07$.

While our expressions for $\Svuln$ and $\Strig$ only
depend coarsely on $\vec{\infprob}_1$, the final extent
of a global spreading event $S$ is more sensitive to
changes in node response functions.  
Taking as an example 
one of our bimodal networks with
$k_a=3$ and $k_b=10$, we see any value of $\phinode$
in $(\frac{1}{10},\frac{1}{3}]$ 
is equivalent to $\phinode=\frac{1}{3}$, as far
as $\Svuln$ and $\Strig$ are concerned,
since only degree 3 nodes are vulnerable.
However, whether or not degree 10 nodes are vulnerable
to $i$ neighbors switching to state $\state{1}$ depends on $\phinode$
meeting and exceeding $\frac{i}{10}$.
For the case $k_a=3$ and $k_b=4$, the results we
present here apply for all $\phinode$ in $(\frac{1}{4},\frac{1}{3}]$.

For the final size of global spreading events, $S$, 
manipulations of \Req{eq:tccn.gleeson_corr_r}
for $k_a=3$ and $k_b=4$
show we need to solve the following for $\theta_{3,\infty}$:
\begin{equation}
  \label{eq:tccn.Sr3/4}
  \theta_{3,\infty} 
  =
  \frac{1+r}{2}i
  (2-\theta_{3,\infty})\theta_{3,\infty} 
  +\frac{1-r}{2}
  (3-\theta_{4,\infty})\theta_{4,\infty}^2
\end{equation}
where
\begin{equation}
  \label{eq:tccn.Sr3/4_b}
  \theta_{4,\infty} = \frac{1}{1-r}\theta_{3,\infty} 
  (1 - 3r + 2r\theta_{3,\infty})
\end{equation}
for $0 \le r < 1$.  Knowing that $\theta_{3,\infty}=1$
is a solution of \Req{eq:tccn.Sr3/4} means we can
reduce the problem to solving for the roots of a quartic, for which
we could obtain a complete analytic solution.  
For our purposes, numerical simulation is sufficient.

Our simulations and numerical analyses indicate the presence 
of another kind of phase transition in $S$. 
At the upper limit of $r=1$, the network
is separated into two giant, fully-connected components,
one comprising solely $k=3$ degree nodes 
and the other $k=4$ degree nodes
(the giant component is of fractional size 1 for
all other values of $r$).
Spreading therefore occurs only in the 
$k=3$ degree node component when $r=1$,
and, as shown in Fig.~\ref{fig:tccn.modelfit}(a),
the three quantities $S$, $\Svuln$,
and $\Strig$ all equal 4/7 for complete
assortativity.

As we decrease $r$ however, the possibility
that a degree $k=4$ node is connected to two
(or more) degree $k=3$ nodes in the giant vulnerable
component increases.  The expected
size $\phi_\infty$ of a global spreading event grows
gradually until it reaches a discontinuous phase
transition and $\phi_\infty$ jumps to 1.
Through numerical analysis of \Req{eq:tccn.Sr3/4}, we see that as
$r$ increases, a repeated root $\theta_{3,\infty}$ appears when
$r = r_c^{\rm upper} \simeq 0.775$.
The roots become real and separate
with both of them limiting to 1 as $r \rightarrow 1$.
Fig.~\ref{fig:tccn.modelfit}(b) shows how both $\theta_{3,\infty}$ and
$\theta_{4,\infty}$ behave as a function of $r$.
Note that the discontinuity in the final size of a global
spreading event is not reflected in the curves
for $\Strig$ or $\Svuln$, which both
change smoothly for $0 \le r \le 1$.

\section{Concluding remarks}
\label{sec:tccn.conclusion}

We have developed a series of theoretical expressions 
for a general class of spreading models on random
networks with tunable degree assortativity $r$.  
Our main results are the derivation of the fractional size of
the giant vulnerable and triggering components as
a function of degree-degree distribution $e_{jk}$.
When we allow all nodes to be susceptible
with uniform probability $\infprob$, our results reduce to the
standard disease-like spreading case, and when
we further set $\infprob=1$, we obtain the conditions
for the existence of a giant component.

When we set $r=0$, we also retrieve the condition for the existence
of a giant vulnerable component (\Req{eq:tccn.F_j'cond}) 
along with its size (\Req{eq:tccn.G1}) in pure
random networks with arbitrary degree distributions~\citep{watts2002a}.
For the triggering component, aside from our work
for general $r$, we have also added to known results for
the $r=0$ case.  We have obtained the probability
that an exogenously activated node of degree $k$ may
trigger a global spreading event (\Req{eq:tccn.Strigk}) and
hence the probability that a randomly chosen node
may do the same (\Req{eq:tccn.H1}).  Our work complements
the results of Gleeson~\citep{gleeson2008a} which determines
the final size of a global spreading event 
(Eqs.~\req{eq:tccn.gleeson_corr_phi} and \req{eq:tccn.gleeson_corr_r}).
We note that a basic description of many kinds of spreading
from a single initial seed must separately report 
(1) the probability of a global spreading event $\Ptrig$, and
(2) the expected final size $S$.  The distribution of final
sizes is often bimodal and an overall expected size conflates
$\Ptrig$ and $S$ in a potentially misleading way, since no
size other than 0 and $S$ are in fact possible.
Finally, the simple family of random networks we have considered here
with $P_k = \frac{4}{7} \delta_{k3} + \frac{3}{7} \delta_{k4}$
have allowed us to demonstrate, for one example,
the validity of our theoretical results for $\Svuln(r)$ and $\Strig(r)$.

\acknowledgments
JLP was supported by a Graduate Research Fellowship 
awarded by Vermont EPSCoR (NSF EPS 0701410).
The authors are grateful for the computational 
resources provided by the Vermont Advanced Computing 
Center which is supported by NASA (NNX 08A096G).

\appendix

\section*{Appendix: Calculations regarding collapse of general global spreading condition for uncorrelated networks}

In the main text, we noted that the general condition for global spreading,
that
$
  \left| \mathbf{A}_{\mathbf{E},\vec{\infprob}_1 } \right| 
  = 0
$
(\Req{eq:tccn.F_j'cond}),
collapses to the appropriate condition for
uncorrelated networks when we set $e_{jk} = q_j q_k$ (\Req{eq:tccn.uncorrcond}).
We demonstrate this in two distinct ways.
First, we take a direct route by setting $e_{jk}=q_j q_k$ in the definition of
$\mathbf{A}_{\mathbf{E},\vec{\infprob}_1 }$:
\begin{eqnarray}
  \label{eq:tccn.Asimp}
  \left[ \mathbf{A}_{\mathbf{E},\vec{\infprob}_1 } \right]_{j+1,k+1}
  & = & 
  \delta_{jk} q_{k} 
  -
  k \infprob_{k+1,1} q_j q_k
  \nonumber \\
  & = & 
  \left(
    \delta_{jk}
    -
    q_j k \infprob_{k+1,1}
  \right)
  q_k.
\end{eqnarray}
The $q_k$ that has been factored out
stands as a multiple for the $k$th column,
and therefore, by multilinearity of determinants~\citep{strang2009a},
contributes to the determinant of $\mathbf{A}_{\mathbf{E},\vec{\infprob}_1}$ 
as a simple overall multiplicative factor.
In finding 
$|\mathbf{A}_{\mathbf{E},\vec{\infprob}_1}|$,
we can therefore focus on the matrix
$
[
\delta_{jk}
-
q_j k \infprob_{k+1,1}
].
$
We find the determinant of this matrix by finding its
eigenvalues, which are the eigenvalues of $[-q_j k \infprob_{k+1,1}]$
incremented by 1 due to the addition of the identity matrix.
Since $[-q_j k \infprob_{k+1,1}]$ is an outer product of
the two vectors $[-q_j]$ and $[k \infprob_{k+1,1}]$, we have
a rank one matrix which therefore has a sole non-zero eigenvalue
corresponding to an eigenvector in the direction of $[-q_j]$.
We find the non-zero eigenvalue by explicitly applying
the matrix $[-q_j k \infprob_{k+1,1}]$ to $[-q_j]$, giving us
$\lambda_1 = -\sum_{k=0}^{\infty} k \infprob_{k+1,1} q_k$.
Shifting the summation index, we have
$\lambda_1 = -\sum_{k=1}^{\infty} (k-1) \infprob_{k,1} q_{k-1}$,
and $\lambda_i = 0$ for $i \ge 2$.
We now add 1 to all eigenvalues to obtain 
$\lambda_1' = 1-\sum_{k=1}^{\infty} (k-1) \infprob_{k,1} q_{k-1}$
and $\lambda_i' = 1$ for $i \ge 2$.  The determinant
of $\mathbf{A}_{\mathbf{E},\vec{\infprob}_1}$ is given by
the product of the $\lambda_i'$ along with the
product of the $q_k$ elements we initially factored out:
\begin{equation}
  \label{eq:tccn.Auncorr}
  |\mathbf{A}_{\mathbf{E},\vec{\infprob}_1}|
  =
  \left(
  \prod_{k=0}^{\infty} q_k
  \right)
  \left(
    1-\sum_{k=1}^{\infty} (k-1) \infprob_{k,1} q_{k-1}
  \right).
\end{equation}
Since the $q_k$ are not equal to 0, the condition
that $|\mathbf{A}_{\mathbf{E},\vec{\infprob}_1}|=0$
for uncorrelated networks depends on the second term:
$0 = 1-\sum_{k=1}^{\infty} \infprob_{k,1} (k-1)  q_{k-1}$.
Substituting $q_k = kp_k/\kavg$ gives us the 
desired condition of~\Req{eq:tccn.uncorrcond}.

Secondly, we can also show that $F_{k}'(1;\vec{\infprob}_1)$ is
independent of $k$, meaning that the expected size of 
a finite subcomponent found at the other end of an edge
connected to a node of degree $k+1$ does not depend
on the latter's degree.  This will also lead us to
the same condition for global spreading we found above.
To do so, we return to 
\Req{eq:tccn.F_j'3} and set $e_{jk} = q_j q_k$:
\begin{equation}
  \label{eq:tccn.F_j'2mod}
  \sum_{k=0}^\infty
  \left(
    \delta_{jk} q_k 
    -
    k \infprob_{k+1,1} q_j q_k
  \right)
  F_k'(1;\vec{\infprob}_1)
  =
  \sum_{k=0}^{\infty}
  q_j q_k
  \infprob_{k+1,1}.
\end{equation}
Breaking the left hand side into two pieces,
carrying out the first summation, and then dividing through by $q_j$,
we obtain
\begin{equation}
  \label{eq:tccn.F_j'2mod2}
  F_j'(1;\vec{\infprob}_1)
  -
  \sum_{k=0}^\infty
  k \infprob_{k+1,1} q_k
  F_k'(1;\vec{\infprob}_1)
  =
  \sum_{k=0}^{\infty}
  q_k
  \infprob_{k+1,1}.
\end{equation}
Upon rearranging, we can now express $F_j'(1;\vec{\infprob}_1)$ as
a function of terms not involving $j$ (even if they include summations
involving $F_k'(1;\vec{\infprob}_1)$), thereby demonstrating
independence:
\begin{equation}
  \label{eq:tccn.F_j'2mod3}
  F_j'(1;\vec{\infprob}_1)
  =
  \sum_{k=0}^\infty
  k \infprob_{k+1,1} q_k
  F_k'(1;\vec{\infprob}_1)
  +
  \sum_{k=0}^{\infty}
  q_k
  \infprob_{k+1,1}.
\end{equation}
Setting $F_k'(1;\vec{\infprob}_1) = F'(1;\vec{\infprob}_1)$ for all $k$, 
the above is now solvable, and we find
\begin{equation}
  \label{eq:tccn.F_j'2mod4}
  F'(1;\vec{\infprob}_1)
  =
  \frac{
    \sum_{k=0}^{\infty}
    q_k
    \infprob_{k+1,1}.
  }
  {
    1-
    \sum_{k=0}^\infty
    k \infprob_{k+1,1} q_k
  }.
\end{equation}
Finally, we see that setting the denominator to 0 recovers
the condition for global spreading in uncorrelated networks.


\begin{thebibliography}{27}
\bibitem[{foo()}]{footnote:tccn.composition}
\bibinfo{note}{The composition rule applies for a sum of identical random
  variables where the number summed is itself a random variable (the emanating
  degree in our particular case). Specifically, if we have two non-negative
  integer variables $N^{(1)}$ and $N^{(2)}$ with probabilities
  $P^{(1)}_{N^{(1)}}$ and $P^{(2)}_{N^{(2)}}$, and generate a third random
  variable as $N^{(3)} = \sum_{l=1}^{N^{(2)}} {N^{(1)}}_l$, then
  $\theta^{(3)}(x;\vec{\infprob}_1) =
  \theta^{(2)}(\theta^{(1)}(x;\vec{\infprob}_1))$ where $\theta^{(i)}$ is the
  generating function for $P^{(k)}_{N^{(i)}}$.}
\end{thebibliography}
\end{document}